%% file: main.tex
\PassOptionsToPackage{table,svgnames}{xcolor}
\documentclass[sigplan]{acmart}
\renewcommand\footnotetextcopyrightpermission[1]{}
\usepackage{color,soul}
\usepackage{subcaption}
\usepackage[linesnumbered, ruled, vlined]{algorithm2e}
\usepackage{multirow}

\usepackage{xcolor}
\usepackage{afterpage}

\usepackage{pifont}

\usepackage{amsmath}

\newcommand{\ProjectName}{\textit{LL-GABR}}

\AtBeginDocument{%
  \providecommand\BibTeX{{%
    \normalfont B\kern-0.5em{\scshape i\kern-0.25em b}\kern-0.8em\TeX}}}

\setcopyright{acmcopyright}
\copyrightyear{2018}
\acmYear{2018}
\acmDOI{XXXXXXX.XXXXXXX}

\acmConference[]{February 14,
  2024}
\acmPrice{15.00}
\acmISBN{978-1-4503-XXXX-X/18/06}

\begin{document}


\title{{\ProjectName}: Energy Efficient Live Video Streaming using Reinforcement Learning}


\author{Adithya Raman}
\email{araman5@buffalo.edu}
\affiliation{
  \institution{University at Buffalo}
  \streetaddress{Davis Hall}
  \city{Buffalo}
  \state{New York}
  \postcode{14260}
  \country{United States}
  }

\author{Bekir Turkkan}
\email{b.turkkan@ibm.com}
\affiliation{%
  \institution{IBM Research}
  \streetaddress{Davis Hall}
  \city{Yorktown}
  \state{New York}
  \postcode{10598}
  \country{United States}
}

\author{Tevfik Kosar}
\email{tkosar@buffalo.edu}
\affiliation{%
  \institution{University at Buffalo}
  \streetaddress{Davis Hall}
  \city{Buffalo}
  \state{New York}
  \postcode{14260}
  \country{United States}
}

\renewcommand{\shortauthors}{Raman.A}

\begin{abstract}
Over the recent years, research and development in adaptive bitrate (ABR) algorithms for live video streaming have been successful in improving users' quality of experience (QoE) by reducing latency to near real-time levels while delivering higher bitrate videos with minimal rebuffering time. However, the QoE models used by these ABR algorithms do not take into account that a large portion of live video streaming clients use mobile devices where a higher bitrate does not necessarily translate into higher perceived quality. Ignoring perceived quality results in playing videos at higher bitrates without a significant increase in perceptual video quality and becomes a burden for battery-constrained mobile devices due to higher energy consumption. In this paper, we propose {\ProjectName} (Low Latency), a deep reinforcement learning approach that models the QoE using perceived video quality instead of bitrate and uses energy consumption along with other metrics like latency, rebuffering events, and smoothness. {\ProjectName} makes no assumptions about the underlying video, environment, or network settings and can operate flexibly on different video titles, each having a different bitrate encoding ladder without additional re-training, unlike existing learning-based ABRs. Trace-driven experimental results show that {\ProjectName} outperforms the state-of-the-art approaches by up to 44\% in terms of perceptual QoE and a 73\% increase in energy efficiency as a result of reducing net energy consumption by 11\%.
\end{abstract}




\maketitle

\section{Introduction}
\label{s:introduction}
\input{introduction}

\section{Background and Related Work}
\label{s:related_work}
\input{related_work}

\section{{\ProjectName} Design}
\input{design}

\section{Evaluation}
\label{s:evaluation}
\input{evaluation}


\section{Conclusion}
\label{s:Conclusion}
\input{Conclusion}

\bibliographystyle{ACM-Reference-Format}
\bibliography{sample-base}

\end{document}

%% file: introduction.tex
According to the recent Ericsson Mobility Report \cite{erricson}, there has been a surge in mobile data traffic, with current trends in 2023 showing 71\% of the global mobile traffic attributed to mobile video streaming with forecasts predicting an increase to 80\% by 2028. This combined with a 3.5x increase in overall mobile data traffic results in the traffic for video streaming to be around 85 Exabytes in 2023 and increasing to 263 Exabytes per month by 2028. The effect of video on global traffic is more pronounced as more video providers start offering HD and UHD (2K and 4K) streaming \cite{ciscoMobileForecast}.

Video Streaming over HTTP has grown tremendously over recent years with the advent of Dynamic Adaptive Streaming over HTTP (DASH) \cite{dashAbout}. In DASH, the video is encoded at multiple bitrates, and the DASH client can use Adaptive Bitrate (ABR) algorithms to select the appropriate bitrate for the next chunk based on network and player buffer conditions. Video streaming can be either in the form of Video-on-Demand (VoD) or Live Streaming. In Video-On-Demand (VoD), where the content is pre-recorded and stored on a server, the ABR algorithm aims to maximize bitrate while minimizing rebuffering events. In Live Streaming, which is the focus of this paper, the content is played on the client's device as it is being captured, leading to an additional constraint on minimizing end-to-end latency to improve the user's Quality of Experience (QoE).

Live video streaming, in particular, has seen tremendous growth in this period due to the (i) pervasive rollout of 5G networks, (ii) the next generation of smartphones being able to produce full HD and 4K UHD videos \cite{ciscoMobileForecast}, and (iii) the increase in popularity of live streaming applications like Twitch which amassed around 24.3 billion hours \cite{twitchTracker} in 2021. An increase in teleconferencing like Zoom and FaceTime has also been significant in the push to deliver better quality low latency live streaming solutions.

Having shorter segment sizes in the range of 1-2 seconds, using Common Media Application Format (CMAF)\cite{cmafEvaluation} and HTTP/1.1 Chunked Transfer-Encoding (CTE), Low-Latency Live (LLL) streaming has undergone rapid development over recent years. The recent Twitch Live Streaming challenge also brought forth several ABR algorithms \cite{lol,l2a, stallion} that can deliver high QoE and low latency streaming under different and varying network conditions.

The current research in ABR algorithms for live streaming, however, does not consider that a significant portion of client devices are mobile phones with smaller screens, and the QoE models designed for the ABR objectives do not consider the perceptual video quality where an increase in bitrate does not necessarily translate to a higher quality video. This further increases energy consumption during streaming, draining the mobile device's battery life.

In this paper, we address these issues faced by LLL by proposing a more realistic QoE evaluation metric and training a Reinforcement Learning (RL) agent that jointly selects both the desired bitrate and speed to maximize the rewards based on the same metric. To this end, we propose {\ProjectName} where we make the following  contributions:

\begin{itemize}
    \item We build an RL-based solution that can learn to adapt both the playback bitrate and playback speed under varying network conditions to maintain high levels of QoE while also simultaneously being more energy-efficient when compared to other SOTA solutions.
    \item We propose a generalized ABR that learns to make intelligent decisions across different bitrate encodings and representation sets without additional training.
    \item We design a modified QoE calculation by replacing bitrate with VMAF values to provide a more accurate evaluation of perceptual video quality. This results in an ABR algorithm that makes bitrate selections based on VMAF scores, where a higher bitrate does not necessarily lead to higher perceptual quality. This also directly translates to lesser rebuffer time and a smoother streaming experience.
    \item We conduct extensive experiments using real-world LTE and 4G/5G traces with realistic bitrate encodings and compare our results against the state-of-the-art ABR algorithms in live streaming.
\end{itemize}

Section \ref{s:related_work} gives an overview of past work in the field of HTTP low latency streaming and ABR algorithms that have been developed for LLL. We also go over some of the existing solutions for energy-efficient HTTP streaming. In Section \ref{s:design}, we introduce the {\ProjectName} model, and Section \ref{s:evaluation} presents the evaluation details and results where we compare our ABR algorithm with the state-of-the-art approaches. Section \ref{s:Conclusion} concludes the paper with a discussion of future work. 

%% file: related_work.tex

\subsection{Adaptive Bitrate (ABR) Streaming over HTTP} 
\label{httpStreaming}

Video streaming over HTTP has become the standard technology to supply the increasing demand of diverse clients. 
Industry leaders and researchers have proposed several algorithms customized for different streaming needs with the development of DASH standards~\cite{dashAbout}. 
Most of the existing studies such as BOLA~\cite{BOLA}, Festive~\cite{festive}, MPC~\cite{mpc}, BOLAE~\cite{dynamicABR}, and Dynamic ABR~\cite{dynamicABR} employ heuristic rules using client buffer level, measured throughput, or both to make bitrate selections for video chunks. 
Despite the practicality of heuristic models, they assume more steady network conditions and commonly fail to adapt to frequent changes.

Several models use machine learning techniques to find optimal decisions under challenging conditions. 
CS2P~\cite{cs2p} and Fugu~\cite{fugu} build supervised learning models with historical data and find optimal bitrate decisions. 
Pensieve~\cite{pensieve}, Comyco~\cite{comyco}, and GreenABR~\cite{greenabr} develop reinforcement learning models to maximize the user's QoE. 
Specifically, Pensieve uses an actor-critic model while Comyco adopts imitation learning with a compute-intensive dynamic programming component, and GreenABR proposes an energy-aware approach with a DQN model.
Despite their competitive performance, all the above algorithms are designed for video-on-demand (VoD) streaming and suffer from large latency and frequent freezing events for live streaming scenarios.

\subsection{Live ABR Streaming}
\label{LLLABRs}

While ABR algorithms for VoD and live streaming share the common goal of providing a high-quality viewing experience to users, out-of-the-box VoD streaming algorithms fall short in achieving near real-time latency requirements of live streaming applications.
VoD services have more flexibility in managing latency and buffering since a large buffer size is available to download pre-recorded video chunks in advance.
In contrast, the near real-time latency constraint of live streaming demands ABR algorithms to use a very small client buffer and to quickly adapt to network fluctuations without causing significant delays or stalling events.



While significant research has been conducted on ABR algorithms for VoD streaming \cite{pensieve,BOLA,comyco,dynamicABR,greenabr}, the progress of ABR for low-latency streaming (LLL) has only recently gained momentum with the first work titled ACTE \cite{bandwidthPrediction}, where Bentaleb \emph{et al.} leveraged a sliding window to measure available bandwidth, a linear adaptive filter to predict future bandwidth and an ABR to select bitrate based on the predicted bandwidth and player buffer levels.

Similar to ACTE, Gutterman \emph{et al.} proposed STALLION \cite{stallion}, which used a sliding window measure to the mean and standard deviation of bandwidth and latency, and built a heuristic rule for the ABR controller to select the appropriate bitrate for playback. Lim \emph{et al.} designed LoL\cite{lol}, which adopted the throughput prediction module in \cite{bandwidthPrediction} and built a learning-based ABR algorithm with Self Organizing Maps (SOM) to turn bitrate selection as an unsupervised classification problem. Karagkioules \emph{et al.} proposed Learn2Adapt-Low Latency (L2A-LL)\cite{l2a}, where the ABR was presented as an optimization framework. The authors used Online Convex Optimization (OCO) to model a learning agent whose objective is to minimize latency with a convex constraint on buffer limit. Unlike \cite{lol,stallion}, in this work bitrate decisions are based on historical values and do not rely on any throughput estimations. To address some of the shortcomings of LoL \cite{lol}, Bentaleb \emph{et al} built LoL+ \cite{lolplus}. Instead of a static set of weights for the SOM, LoL+ utilized a heuristic-based dynamic weight assignment algorithm that adjusted feature weights at the segment download boundary. The paper also used the different metrics of the linear QoE function as individual features for the SOM model.




Sun \emph{et al.} \cite{tightrope}, considered video rate and playback speed as a joint adaptation strategy and developed a Reinforcement Learning based ABR solution using Branching Dueling Q-Networks (BDQ). The BDQ ABR agent was used to select both bitrate and playback speed at every segment boundary to maximize the user's QoE. The BDQ agent had an expanded playback control set and also had the ability to skip/repeat segments that were absent in \cite{lol, stallion,l2a}.

\subsection{Energy Aware ABR Streaming}
\label{ss:energyAwareABR}
Existing video streaming studies consider energy efficiency and QoE as conflicting goals and simply target making optimal decisions for one of them.
They commonly target maximizing QoE and ignore the energy consumption impact~\cite{pensieve,dynamicABR, mpc,BOLA, festive}.
Early energy-efficient streaming efforts focused on hardware-related solutions such as altering screen brightness dynamically~\cite{wowmomPower} or switching off network connections for idle time~\cite{eff-has}. 
Similarly, Chen et al.~\cite{powerModel_icdcs} propose adjusting video quality depending on the video streaming context.
However, they are either tailored for specific use cases or require hardware support. 
Recently, GreenABR~\cite{greenabr} has proposed optimizing QoE and energy efficiency together by exploiting perceptual quality metric, VMAF~\cite{NetFlix2}. It is designed for VoD streaming scenarios and does not consider low-latency requirements.

\subsection{Limitations of Existing Solutions}
\label{ss:limitations}
The LLL ABR algorithms covered in previous sections share some common limitations regarding perceptual quality considerations, energy efficiency, evaluations, and generalizability. 

\begin{itemize}
    \item \textbf{Perceptual Quality:} The QoE model used in most of the existing studies~\cite{lol,l2a,lolplus, stallion, tightrope, pensieve, dynamicABR} fall short in reflecting users' perceived QoE. They assume a linear relation between video quality and encoding bitrate, which is not necessarily true, especially for mobile users~\cite{VMAF-Phone, greenabr}.
    
    \item \textbf{Energy Efficiency:} As a result of not considering the perceptual quality and using a smaller and unrealistic bitrate encoding, the ABR algorithm ends up targeting higher bitrates even if it does not lead to a higher QoE. This adds up to significant energy consumption on battery-constrained mobile devices~\cite{greenabr}.
    
    \item \textbf{Evaluations:} Improvements in mobile devices, networking, and broadband technologies enabled high-quality live streaming. For instance, Twitch player supports 6000 Kbps encoding bitrate for 1080p resolution with 60 FPS~\cite{twitchEncoding}, and Netflix recommends above 4000 Kbps for 1080p videos~\cite{ntflixRecommendedSet}. However, several existing studies, including LoL~\cite{lol}, L2A-LL~\cite{l2a}, and LoL+~\cite{lolplus} use a narrower bitrate range, 1000 Kbps for 720p resolution, for their evaluations. 


    \item \textbf{Lack of generalizability:} Modern streaming applications use large representation sets for their videos due to client diversity.
    It is especially important for learning-based models to generalize over different representation sets. 
    They commonly require additional training to be used for different bitrate encoding ladders~\cite{pensieve, greenabr, comyco, tightrope}, which cause more computing costs and unstable performance.
    
    
\end{itemize}

In this work, we address the above limitations and propose Green-LL, which optimizes bitrate selections for both energy efficiency and QoE. It supports a large encoding bitrate range without requiring additional training or tuning parameters for each representation set.

%% file: design.tex
\label{s:design}


In this section, we will introduce the design methodology for {\ProjectName}. We talk about our design goals (Section \ref{ss:design_goals}) and differences in live streaming vs VoD streaming. We also briefly touch upon the energy model (Section \ref{ss:power_model}) used to estimate energy consumption for streaming live video and the bandwidth prediction(Section \ref{ss:bw_prediction}) module for prediction of the bandwidth of the next video chunk. Finally, we end this section by discussing the Soft-Actor-Critic algorithm (Section \ref{ss:rl}) in detail, where we also cover why an RL-based algorithm is beneficial and why we use a continuous action space-based algorithm. 

\subsection{Design Goals}
\label{ss:design_goals}
{\ProjectName}'s fundamental goal is to ensure energy-efficient live video streaming on mobile devices. This includes reducing the total energy consumption of streaming while maintaining a high QoE. We define the QoE in Section \ref{ss:qoe} to include various player metrics like perceptual video quality, latency, freezing, smoothness, and playback rate. We design {\ProjectName} to consider the VMAF values of every video chunk to evaluate the quality of the video segment and also the energy consumption of each chunk. This is done by incorporating both these metrics in the reward function for the RL agent, which is detailed in Section \ref{ss:reward}.

We also address the limitations of existing ABRs which were designed with only a single constant bitrate ladder in mind. {\ProjectName} is built to be more of a one-size-fits-all model where an RL agent trained once can adapt to different video titles, each having their own bitrate ladders. This is particularly advantageous as streaming providers begin to adopt per-title encoding for live video workflows, where video content determines the encoding settings for each video title.  

In designing {\ProjectName}, we must note a major difference between traditional VoD and live streaming, which is the additional constraint on latency and buffer levels. In Live Streaming, we measure the glass-to-glass latency, the time difference between when the content was captured/generated at the source and viewed at the client's device. In VoD, since the content has already been created and stored on a server for the viewer to fetch, measuring the glass-to-glass latency is less important. As a result, instead of having only bitrate in the action space, in live streaming, we also have to control the playback rate (speed) with which the content is played. A higher playback rate leads to a lower latency, while a lower playback rate leads to more buffer levels and higher latency.

\subsection{Bandwidth Prediction}
\label{ss:bw_prediction}
Accurate bandwidth prediction is critically important in low latency ABR as it helps in selecting the appropriate bitrate for streaming. If bandwidth is overestimated, the chosen bitrate may be too high, leading to rebuffering. Conversely, underestimating the bandwidth might result in a lower video quality being selected. Some learning-based ABRs \cite{tightrope} used an LSTM network to predict future bandwidth values. However, this increases the complexity of the RL model leading to higher training costs and time.

In {\ProjectName}, we use the approach proposed by A.Bentaleb \emph{et al.} \cite{bandwidthPrediction}, ACTE, which uses a chunk-based sliding window moving average for bandwidth measurement and an online adaptive filter based on a recursive least square (RLS) approach for bandwidth prediction. This approach is preferred over the LSTM method as it is computationally lighter and can be run on mobile devices without any additional costs.

\subsection{Energy Model}
\label{ss:power_model}
In our experiments, we use an energy model very similar to the one proposed by Turkkan \emph{et al.} \cite{greenabr}. This model computes the local playback and data acquisition energy for every downloaded segment. The \emph{local playback energy} corresponds to the processing and displaying of the video file components, while the \emph{data acquisition} component corresponds to the download of the video files via the wireless network. By discounting the device's base energy consumption, the model also remains device agnostic. 

\subsubsection{\textbf{Local Playback Energy:}}
This includes the energy consumption involved in the decoding, processing, rendering, and displaying of the downloaded chunk. The factors influencing this are video resolution, bitrate, framerate, chunk size, chunk frequency, video complexity, motion rate, and codec. We build a linear regression model trained on collected data to accurately predict the measured playback energy. During streaming, this model is used to calculate the energy consumption of each video chunk.

\subsubsection{\textbf{Data Acquisition Energy:}}
This refers to energy consumption for downloading the video segments over a wireless network, which is related to throughput and file size and is computed using a formula derived from an existing throughput-based model \cite{wifiPowerModel}.

\begin{align}
\label{eq:networkPower}
  E_{data} := \sum_{i=0}^\mathcal{C} \left[\alpha * th_{c_i}^{-1} + \beta * {fs_{c_i}} \right]
\end{align}
$\mathcal{C}$ refers to the number of chunks per segment, $th_{c_i}$ and $fs_{c_i}$ are the predicted bandwidth \cite{bandwidthPrediction} and chunk size of the $i$-th chunk respectively

The local playback energy and the data acquisition energy are summed up to obtain the total energy for video playback for every segment. We refer the reader to the energy model in \cite{greenabr} for more details on the experimental analysis involving how the energy consumption for both of these components is calculated.

\begin{figure*}[t]
\centering

\includegraphics[width=1\linewidth]{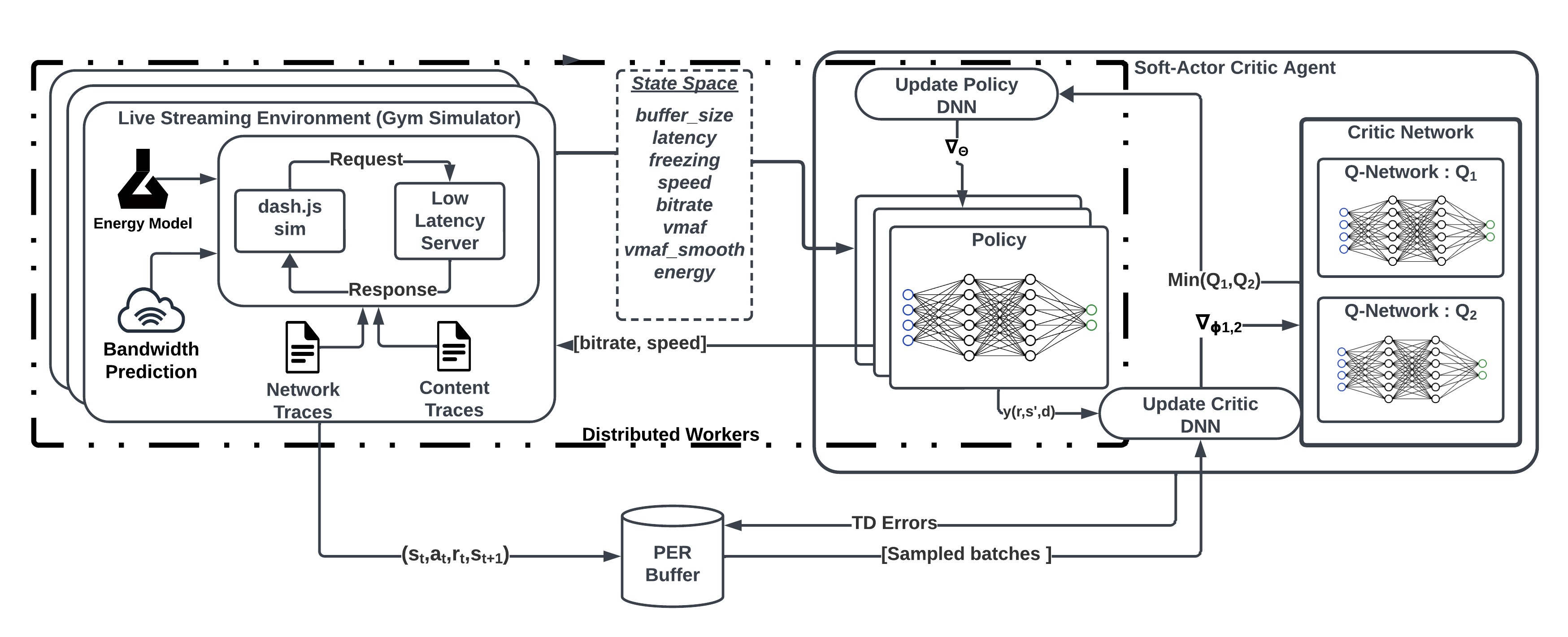}
\caption{Overview of {\ProjectName}'s distributed SAC training.}
\label{fig:sac_overview}
\end{figure*}

\subsection{Reinforcement Learning for Live Streaming}
\label{ss:rl}
\subsubsection{\textbf{Why RL?:}}Reinforcement Learning (RL) has shown considerable success in control tasks for optimization problems \cite{dataCentreCooling} as it offers a dynamic and adaptive approach often superior to traditional heuristic methods. Since RL makes no assumptions about the underlying environment and can learn from direct interaction with the environment, it can learn novel strategies that can adapt over time. This is particularly advantageous in high-dimensional complex environments. Using a customizable reward function, the RL agent can discover certain goal-oriented behaviors without explicit programming, allowing such algorithms to outperform heuristic methods. This holds especially true in a live-streaming environment where the optimal decision-making policy is difficult to formulate or when the environment is subject to unpredictable network changes.

RL can be effectively used for Adaptive Bitrate (ABR) streaming in live video to optimize the quality of the streaming experience. ABR algorithms aim to adapt the video quality based on the viewer's network conditions, such as bandwidth, latency, and packet loss. RL can be utilized to create an intelligent ABR algorithm that learns to make better decisions about the video bitrate, providing smooth playback and optimal quality. The agent can continuously learn and adapt the streaming strategy based on environmental feedback. This is crucial in network conditions that are inherently variable and unpredictable. RL can also take into account the long-term QoE by considering the consequences of current actions and the state of the environment. This means the agent can avoid actions that lead to rebuffering or quality switches. As such, the RL agent can learn an optimal policy through interaction with the environment, regardless of the underlying complexity. This can be extended further by using a distributed RL where several agents can learn by acting on similar environments with different network conditions to improve the learning time and efficiency.

In the context of ABR, the RL problem can be framed as a Markov Decision Process (MDP). The MDP consists of states, actions, rewards, and a transition function. For ABR, the states can include network conditions (e.g., available bandwidth, latency, buffer occupancy), video encoding parameters (e.g., resolution, bitrate, frame rate), and user experience metrics (e.g., average quality, rebuffering events). Actions correspond to the choice of the video bitrate, and rewards can be designed to encourage high video quality, low rebuffering, and minimal bitrate switches.

Additionally, The RL agent can leverage playback speed and bitrate to optimize QoE by minimizing buffering, maintaining video quality, and synchronizing the stream to real-time events. Maintaining a balance between playback speed and buffer occupancy is crucial. The RL agent can observe the current buffer level and adjust the playback speed to prevent buffer underflow (rebuffering) or overflow (increased latency). The agent can increase playback speed when the buffer is full or approaching overflow, and decrease it when the buffer is low or approaching underflow. This adjustment helps maintain low latency while minimizing rebuffering events, thus improving QoE.


As a high-level overview, we present the design of {\ProjectName}'s training phase in Figure \ref{fig:sac_overview}. We use a Python-based OpenAI Gym compatible live streaming simulator to help train our agent. The simulator loads the network and content traces files as inputs and runs a 300-second streaming session for every episode. Since we use a segment size of 1 second and a chunk size of 200\textit{ms}, we have 300 steps for every episode.  In each step, the environment takes an action submitted by the RL agent and outputs a state vector consisting of various player and network metrics. The bandwidth prediction (Section \ref{ss:bw_prediction}) and energy model (Section \ref{ss:power_model}) are included as part of the environment and help in generating predicted bandwidth and consumed energy, respectively, which are appended to the state vector. During the training phase, every experience tuple $(s_t,a_t,r_t,s_{t+1})$ is added to the buffer, which is used to make updates to the Policy and Critic networks of the RL agent. We also use multiple parallel workers, to speed up the training phase.

\subsubsection{\textbf{State Space:}}
At every video segment boundary, $t$, the {\ProjectName} agent reads the player metrics, network dynamics, and player energy consumption as the input state space, $s_t$, for the RL model. The player metrics consist of current buffer size, latency, rebuffering duration, playback rate (speed), bitrate, VMAF, and VMAF smoothness of the current segment. Network dynamics includes the predicted bandwidth value\cite{bandwidthPrediction}. The player's energy consumption comprises the network and playback-decoding energy consumption, as detailed in Section \ref{ss:power_model}. The entire state space is also passed through a MinMax normalizer to bring the values to an equal distribution of (0,1).

\subsubsection{\textbf{Action Space:}}
{\ProjectName}'s policy network processes the input state to yield a two-dimensional continuous action vector, where each element ranges from 0 to 1. The first dimension controls the bitrate which is modeled for a flexible representation set compatible with a per-title encoding schema. This enables the model to adapt its bitrate decisions to various different content with distinct bitrate ladders. This is achieved through a continuous-to-discrete mapping that aligns the floating-point output with the nearest available discrete option.

For the second dimension, the playback rate, the action corresponds to speeds varying from 0.90x to 1.1x, with the default speed set at 1.0x. This allows the model to adjust playback speed in response to latency conditions; accelerating playback above the default to reduce latency and decelerating to prevent buffer depletion and potential stream freezing. If there are sufficient buffer levels, a maximum speed setting of 1.1x enables the video player to compensate for one second of latency in ten seconds.

\subsubsection{\textbf{Priority Experience Replay:}} We also use a Priority Experience Replay (PER) buffer\cite{per} for collecting and sampling experiences learned during training. In the traditional replay buffer used in RL, experiences are usually sampled uniformly at random. However, not all experiences are of equal importance in terms of what the agent can learn. Some experiences might be rarer than others while others might be more frequent but less informative. For instance, in ABR for a particular streaming session, a drop in estimated bandwidth which would result in a rebuffering session, could be considered more important as a learning experience than one where the bandwidth remains the same for a long period of time. PER prioritizes the replay of experiences from which the agent can learn the most by assigning a priority proportional to the estimated temporal difference (TD) error of each experience in the buffer. This encourages more efficient learning by replaying surprising experiences more often.

\subsubsection{\textbf{Distributed Learning Framework:}} On top of the SAC with PER buffer, we speed up training time and efficiency by using multiple parallel actors collecting experiences from different environments, each running on different cores in the computing cluster. The APE-X framework \cite{horgan2018distributed} implemented in RL-Lib uses multiple actors distributed across workers to scale training from a single core to multiple cores. We also enable each worker to run a live streaming environment with different levels of network traces to simulate experiences with differing characteristics. For instance, one worker would run a live streaming session with an average bandwidth of 0.5 MBps while a second worker would run a similar session but with an average bandwidth of 5.8 MBps. This framework of a centralized learner with a critic network and multiple workers with actor networks vastly improves training time and also distribution of experiences collected by each worker. 

\begin{equation}
\label{eq:reward}
    \begin{aligned}
        reward_t =  \frac{k_1 \cdot g_1(vmaf_t) }{1 + k_2 \cdot g_2(total\_energy_t)} \\ 
              - k_3 \cdot  g_3(is\_freezing ) \\
              - k_4 \cdot g_4(freezing\_duration) \\
              - k_5 \cdot g_5( latency) \\
              - k_6 \cdot g_6(speed_t) \\
              - k_7 \cdot g_7(vmaf_t, vmaf_{t-1})
    \end{aligned}
\end{equation}

\subsubsection{\textbf{Reward Model:}}
\label{ss:reward}
The reward model is central to training an agent to learn an optimal policy of low latency, high perceptual quality, minimal rebuffering, and smoothness. Since in this work, we are also exploring the energy efficiency of video streaming, we include a penalty for the energy consumption per segment. These conflicting goals in the reward model can be captured by a linear weighted sum of individual components as shown in Equation \ref{eq:reward} where $k_i$ is a constant that be adjusted depending on the QoE goals for streaming. The reward is calculated per segment of streaming once the last chunk in the segment is downloaded.

The first component gives us a positive contribution to the overall video quality per unit consumption of energy. A penalization for the number of freezing occurrences and total freezing duration is added in the second and third components of the equation. The subsequent two terms capture the penalty applied for high latency and un-normal speed while the last term applies the penalty for fluctuating video quality. This model suggests a system where the highest rewards are given for high video quality per unit energy consumption, minimal freezing and latency, stable playback speed, and consistent video quality over time. The rewards are further normalized to stabilize the learning process. 

\begin{algorithm}
    \footnotesize
    \SetAlgoLined
    Input: Initialize policy parameters $\theta$, $Q$-function parameters $\phi_{1}$, $\phi_{2}$, empty PER buffer $\mathcal{D}$
    
    Set target parameters equal to main parameters $\phi_{targ,1} \leftarrow \phi_{1}, \phi_{targ,2} \leftarrow \phi_{2}$

    \For {\textbf{Each Training Step}}{
        \For{\textbf{Each Distributed Worker in $j$-th environment}}
        {
            Observe state $s_j$ and selection action $a_j \sim \pi_{\theta}(\cdot | s_j)$ \;
            Execute $a_j$ in the environment $j$\;
            Observe next state $s'_j$, reward $r_j$, and done signal $d_j$ to indicate whether $s'_j$ is terminal\;
            Store $(s_j, a_j ,r_j, s'_j, d_j)$ in the PER buffer $\mathcal{D}$\;
            if $s'_j$ is terminal, reset the environment state\;
        }
        
        \If{it's time to update}{
            \For{ $j$ in range number of updates}{
                Randomly sample batch , $B = {(s, a ,r, s', d)}$ from $\mathcal{D}$ \;
                Compute targets for Q functions\;
                $\begin{aligned}
                    && y(r,s',d) = r + \gamma(1-d)  ( \underset{i=1,2}{min} \ Q_{\phi_{targ,i}}(s',a') - \\
                    && \alpha \log \pi_{\theta} (a | s') ) 
                \end{aligned}$\\

                Compute TD-Errors for sampled batch:
                $\begin{aligned}
                    && td\_error = |y(r,s',d) -  \mathcal{Q}_{\phi_{i}}(s,a) |                    
                \end{aligned}$\\

                Update priority values of the sampled batch in the PER buffer $\mathcal{D}$ with $td\_error$
    
                Update Q-functions by one step of gradient descent
                $\begin{aligned}
                && \nabla_{\phi_{i}} \frac{1}{|B|} \sum_{(s, a ,r, s', d) \in B}( \mathcal{Q}_{\phi_{i}}(s,a) -y(r, s',d) )^{2} \\ 
                && \text{for $i=1,2$}
                \end{aligned}$\\

                Update the policy by one step of gradient ascent
                $\begin{aligned}
                    && \nabla_{\theta} \frac{1}{|B|} \sum_{s \in B} ((\underset{i=1,2}{min} \ Q_{\phi_{i}}(s',a'_{\theta}(s)) \\
                    && - \alpha \log \pi_{\theta} (a_{\theta}(s) | s) )
                \end{aligned}$\\ where $a_{\theta}(s)$ is a sample from $\pi_{\theta}(\cdot | s)$ which is differentiable w.r.t $\theta$ \;
                Update target networks with
                $\begin{aligned}
                    &&\phi_{targ,i} \leftarrow \rho\phi_{targ,i} + (1-\rho)\phi_{i}&&& \text{for $i=1,2$}
                \end{aligned} $\\
                
            }
        }  
    }
    \caption{Distributed SAC with Priority Experience Replay Buffer} 
    \label{alg:sac}
\end{algorithm}

\subsubsection{Soft-Actor-Critic}
\label{ss:sac}
For the {\ProjectName} architecture, we use a Soft-Actor-Critic (SAC) \cite{sac} based RL algorithm with a Priority-Experience Replay (PER) buffer \cite{per}. SAC is an off-policy algorithm that optimizes a stochastic policy similar to Deep Deterministic Policy Gradient (DDPG) \cite{ddpg}. However, one key difference is that SAC uses an entropy-regularized framework, which encourages the policy to explore more while exploiting what it has learned with high rewards. This prevents the policy from converging prematurely to a bad local optimum, which is a major drawback of DDPG when applied to a problem with large action space.

SAC incorporates entropy in Equation \ref{eq:entropy} into the reward function, $R(s_t,a_t,s_{t+1})$, updating the RL problem as shown in Equation \ref{eq:policy} where $\pi^{*}$ refers to the best optimal policy for the agent.
\begin{equation}
    \pi^{*} = \underset{\pi}{argmax}\underset{\tau\sim\pi}{E}\left[\sum_{t-0}^{\infty} \gamma^t \left(R(s_t,a_t,s_{t+1}) + \alpha H(\pi(\cdot | s_t\right)\right]
    \label{eq:policy}
\end{equation}

A higher entropy during training encourages more randomness in training. The entropy $H$ of a policy $\pi(\cdot |s_t)$ is computed using the log probability distribution, as defined in Equation \ref{eq:entropy}.
\begin{equation}
    \label{eq:entropy}
    H(\pi(\cdot | s_t)) = \underset{x~\sim P}{E} \left[ -\log \pi(\cdot |s_t)\right]
\end{equation}

The 
$Q$-function in Equation \ref{eq:qfunction} estimates the expected return of taking a particular action in a state. In SAC, we use two separate $Q$-functions to reduce positive bias in the learning step due to the overestimation of $Q$-values. SAC concurrently learns the policy $\pi_\theta$ and two $Q$-functions $Q_{\phi_1},Q_{\phi_2}$.
\begin{equation}
    \label{eq:qfunction}
    Q^{\pi}(s, a) = E_{\tau \sim \pi} \left[
    \sum_{t=0}^{\infty} \gamma^t R(s_t, a_t, s_{t+1}) + \alpha \sum_{t=1}^{\infty} \gamma^t H(\pi(\cdot | s_t))
    \right]
\end{equation}

%% file: evaluation.tex

\subsection{Network Traces}

In the proposed system, each segment encompasses one second of video, while each chunk spans 200 milliseconds. Simulated live sessions have a duration of 300 seconds. We use a collection of 3G, LTE, 4G and 5G traces \cite{3gtraces,ltenyu,4gtraces,LTEtraces,lumos} for training and testing. Although 3G networks are not as common as 4G and LTE, they are useful in training to help the agent understand low-bandwidth behaviors and hence used only in training but not testing simulations. We also build a collection of synthetic traces that have high degrees of fluctuation for use during testing.

\begin{table}[H]
    \centering
    \rowcolors{3}{gray!25}{white}
    \begin{tabular}{*4c}
        \hline
        \multirow{2}{*}{\textbf{Trace}} & \multirow{1}{*}{\textbf{Avg b/w}} & \multirow{2}{*}{\textbf{\# > 12Mbps}} & \multirow{2}{*}{\textbf{Dist \%}}  \\
        & \textbf{(Mbps)} &  & 
        \\    
        \hline
            3G/HSDPA          & 0.82&0\% &20\%  \\ 
            NYU LTE          & 3.86&0\% &20\%  \\ 
            4G LTE         & 30.21 &90\%  &10\% \\
            5G Lumous         & 520.66 &100\%  &30\%  \\
            Synthetic   & 3.07  &0\%  &20\%  \\
        \hline   
    \end{tabular}
    \caption{Distribution of Network Traces}
    \label{tab:trace_distribution}
\end{table}

\vspace{-9mm} 
\subsection{Video Content Traces}
To train our RL agent and evaluate its performance, we utilize three distinct video datasets \cite{xiph} from different genres, each varying in terms of scene complexity and motion rate and hence encoded with different bitrate ladders. An Animation (\textit{Big-Buck-Bunny}) has less scene variability in terms of texture and motion rate and hence has a ladder that is adequately represented at lower bitrates. We use \textit{Tears-Of-Steel} for the live-action movie genre as it characterized by higher complexity than BBB and has an encoding scheme to accommodate higher bitrates. Finally, we have a Sports (\textit{Football}) video which is more complex in terms of motion rate than the other two with faster and unpredictable scene movement. Such a video would require a much higher set of bitrate representations. The details of resolution-bitrate mappings for each of these videos are presented in Table \ref{tab:video_encoding}. All videos were encoded using \textit{H.264}.

\begin{table}[H]
    \centering
     \rowcolors{4}{gray!25}{white}
    \begin{tabular}{*4c}
    \toprule
        \multirow{2}{*}{\textbf{Resolution}} & \multicolumn{3}{c}{\textbf{Bitrate (Kbps)}} \\ 
            \cmidrule(lr){2-4}
         & \textbf{Animation}  & \textbf{Movie}  & \textbf{Sports} \\ 
        \midrule
        144p & 300 & 375 & 450 \\
        
        360p & 450 & 560 & 670 \\
        360p & 600 & 750 & 900 \\
        
        480p & 800 & 1050 & 1250 \\
        480p & 1400 & 1750 & 2100 \\
        
        720p & 1900 & 2350 & 2800 \\
        720p & 2400 & 3000 & 3600 \\
        
        1080p & 3500 & 4300 & 5200 \\
        1080p & 4700 & 5800 & 7000 \\
        
        2K & 5000 & 8000 & 9000 \\
        2K & 8000 & 10000 & 12000 \\
        \bottomrule
        
    \end{tabular}
    \caption{Video Content-Encoding Scheme}
    \label{tab:video_encoding}
\end{table}
\vspace{-10mm} 
\subsection{Quality of Experience}
\label{ss:qoe}

The QoE of video streaming is an objective metric of the user's streaming satisfaction rating. This can be sufficiently quantified using Equation \ref{eq:qoe}. While the reward function in Equation \ref{eq:reward} provides a numerical signal at each step indicating how benificial or detrimental an action taken by agent is towards achieving its goal, the QoE is a measure of the user's overall satisfaction with the streaming for the entire viewing experience. However it should be noted that the reward function is typically aligned with improving QoE where action's that lead to higher video quality and lower latency would be rewarded more positively as they more likely to enhance the QoE. We divide VMAF value by 20 to get the video quality level (0-5). This QoE formula is similar to the Linear QoE \cite{greenabr} with the added latency and speed penalties. For our experiments, we use $\alpha=0.077$, $\beta=1.249$, $\gamma=2.897$, $\sigma=1.249$, $\mu=0.771$ and $\omega=1.436$

\begin{equation}
    \begin{aligned}
        QoE = \alpha \cdot \sum_{t=0}^T \left( \frac{vmaf_t}{20} \right) - \beta \cdot \sum_{t=0}^T rt_t - \gamma \cdot r_c \\
         - \sigma \cdot \frac{\sum_{t=0}^T l_t}{T} - \mu \cdot |1 - s_t| \\
         - \omega \cdot \sum_{t=2}{T} \left( \frac{|vmaf_t - vmaf_{t-1}|}{20} \right)
    \end{aligned}
    \label{eq:qoe}
\end{equation}

Additionally, to calculate the energy efficiency of video streaming, we incorporate energy consumption into our QoE metric to derive a measure of QoE per unit of energy, denoted as QoE/Energy. This metric is calculated by dividing the total QoE by the total energy consumed during the streaming session. This energy efficiency metric enables us to balance the trade-off between providing a high-quality streaming experience and maintaining energy conservation, which is particularly important for mobile devices.

\subsection{Evaluation Results}
\label{s:results}
\input{result}

%% file: result.tex
\input{mega_table_3}

\begin{figure*}[t]
\centering
\includegraphics[width=1.0\linewidth]{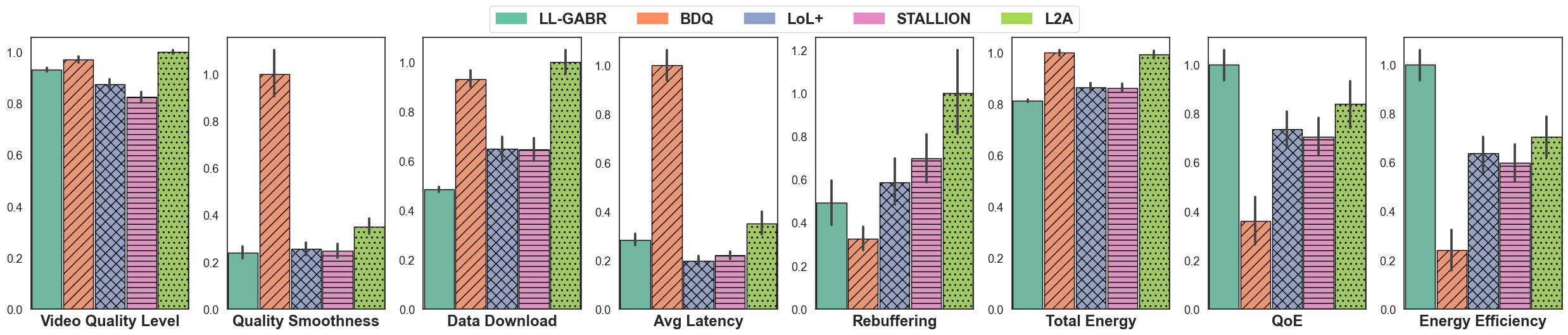}
\caption{Testing results averages on all traces across all videos.}
\label{fig:all_results}
\end{figure*}

 In Figure \ref{fig:all_results}, we present the normalized values for metrics such as \textit{QoE, total energy consumption, energy efficiency, data download volume, video quality, average latency,} and \textit{total rebuffering duration}. To facilitate comparative analysis, we scale the highest value in each category to a score of one, providing a view of the relative performance of {\ProjectName} against other ABR algorithms. These results are compiled through simulations that apply the testing traces detailed in Table \ref{tab:trace_distribution} and the video encodings specified in Table \ref{tab:video_encoding}. Notably, {\ProjectName} undergoes training exclusively with the \textit{Animation} video but is tested across all three video datasets. In contrast, the BDQ model requires individual training sessions for each video. All tests are conducted within a Python 3-enabled trace-driven live-streaming simulator. 
 
In Section \ref{ss:overall_analysis} we go over the different metrics we are interested in and compare the performance of {\ProjectName} with the other baseline ABRs. We shall also look at how {\ProjectName} manages fluctuating throughput conditions in Section \ref{ss:individual_trace}. In Section \ref{ss:training_perf} and Section \ref{ss:cost}, we shall cover the training performance evaluation and how a distributed Soft-Actor-Critic network speeds up the training time and efficiency which we compare with the other RL-based ABR i.e. BDQ. We also briefly talk about the cost in terms of CPU percentage for the training process and the energy cost of the ABR model itself. Both {\ProjectName} and \textit{BDQ} are run three times with different seed settings and the average result is taken.

\subsubsection{\textit{\textbf{Testing Simulation Results}}}
\label{ss:overall_analysis}
We present Table \ref{tab:all_results} which details the performance metrics for each ABR algorithm across all three video contents, providing insights into how each algorithm's effectiveness varies with different types of videos. When we look at the video quality level, QoE, and energy efficiency, the higher the better while for bitrate, data download, latency, freezing, and energy consumption we prefer lower values. For speed, the closer the value is to default (1.0x), the better the score.


\textbf{Quality, Bitrate \& Smoothness:} We measure video quality as a function of VMAF. In Table \ref{tab:all_results}, we note that {\ProjectName} has sufficiently high levels of video quality comparable to the other baselines, only $1\%-2\%$ less than the highest in the \textit{Animation} and \textit{Sports}. In the \textit{Movie} video, the drop in quality is around $8\%$ when compared to BDQ and L2A, but we see that BDQ suffers from higher latency due to poor speed control and L2A has $50\%$ higher rebuffering duration among all the ABRs. We also note that {\ProjectName} is able to achieve these comparable (if not high) levels of perceptual quality even though it has the lowest average bitrate ($\downarrow48\%)$. This is due to the agent selecting bitrate decisions based the VMAF values in order to achieve a high perceptual quality. 

It is also observed that {\ProjectName} has the lowest (best) quality smoothness in the Animation video, while it has the second lowest values in the Movie and Sports videos where Stallion has the best scores.

\textbf{Data Usage:} We also note that our algorithm manages to have the lowest data usage ($\downarrow 35\%$) while maintaining these high VMAF values. This can again be attributed to the {\ProjectName}'s strategy of selecting bitrate actions based on VMAF and quality levels, resulting in high-quality video comparable to other benchmarks without high data usage.

\textbf{Freezing \& Latency:} When we look at freezing and latency, we see that {\ProjectName} performance is comparable to other ABRs. There is a less than one-second difference in latency when compared with LoL+, which is the current state-of-the-art for low latency live streaming ABR. BDQ has the lowest freezing values for ToS and Sports videos, but this is the result of the agent always selecting the lowest playback rate (0.90x), resulting in high buffer levels and consequently leading to a $184\%$ increase in latency. {\ProjectName} is able to balance buffer levels better across all video datasets, maintaining both low freezing and latency rates.

\textbf{Energy Consumption:} By integrating VMAF values (instead of bitrate) and energy consumption into the reward function of {\ProjectName}, we see that the agent has learned behavior where it makes bitrate decisions based on the quality levels, i.e., it does not select the highest bitrate if it does not significantly improve perceptual quality. As a result of using lower bitrates and consequently lower data download, {\ProjectName} has the lowest energy consumption ($\downarrow 11\%$) among all the ABRs. This is due to decreased data acquisition and local playback energy.

\textbf{QoE \& Energy Efficiency:} In Section \ref{ss:qoe}, we define QoE as a linear weighted combination of video quality, rebuffering, freezing, smoothness, and speed. We also divide the QoE with the normalized value of total energy consumption to obtain the QoE per unit energy consumption (or Energy Efficiency). For both these metrics we see that {\ProjectName}, outperforms all the other ABRs with an average of $44\%$ increase in QoE and $73\%$ increase in energy efficiency. 

BDQ in particular suffers from a very low QoE especially for the Movie and Sports videos while it performed comparatively better on Animation. This is because the same training hyperparameters used in Animation are used for the Movie and Sports videos. Individual hyperparameter tuning and training for each title might improve the overall scores and performance of BDQ but this is not a practical solution for any video service provider which has several hundreds of video titles.

\subsubsection{\textbf{\textit{Simulation on individual traces:}}}
\label{ss:individual_trace}

    
    

\begin{table*}[h]
\centering

\rowcolors{3}{gray!25}{white}
\begin{tabular}{*8c}
\hline
\multirow{2}{*}{\textbf{ABR}} & \multirow{1}{*}{\textbf{Training}} & \multirow{1}{*}{\textbf{\#Episodes}} & \multirow{2}{*}{\textbf{CPU\%}}& \multirow{2}{*}{\textbf{\#Processes}} &\multirow{2}{*}{\textbf{\#Cores}} & \multirow{1}{*}{\textbf{Memory Use}} & \multirow{1}{*}{\textbf{Energy}} \\
& \textbf{Time (hrs)} &(for convergence) & && & (in Gb) & (in Kj)\\
\hline

SAC-1  & 1.80 & 5000 & 88\% & 1 &2 & 3.23 & 105\\

SAC-3 & 1.25 & 1666 & 240\% &3 & 4 & 3.23 &128 \\

SAC-5 & 0.80 & 1000 & 300\% &5 & 6 & 3.23 & 156\\

BDQ & 7.80 & 50000 & 122\% &1 & 2 & 3.98 &900\\

\end{tabular}
\caption{Training Cost Metrics}
\label{tab:cost}
\end{table*}

\begin{figure}[ht]
\centering
\includegraphics[width=1\linewidth]{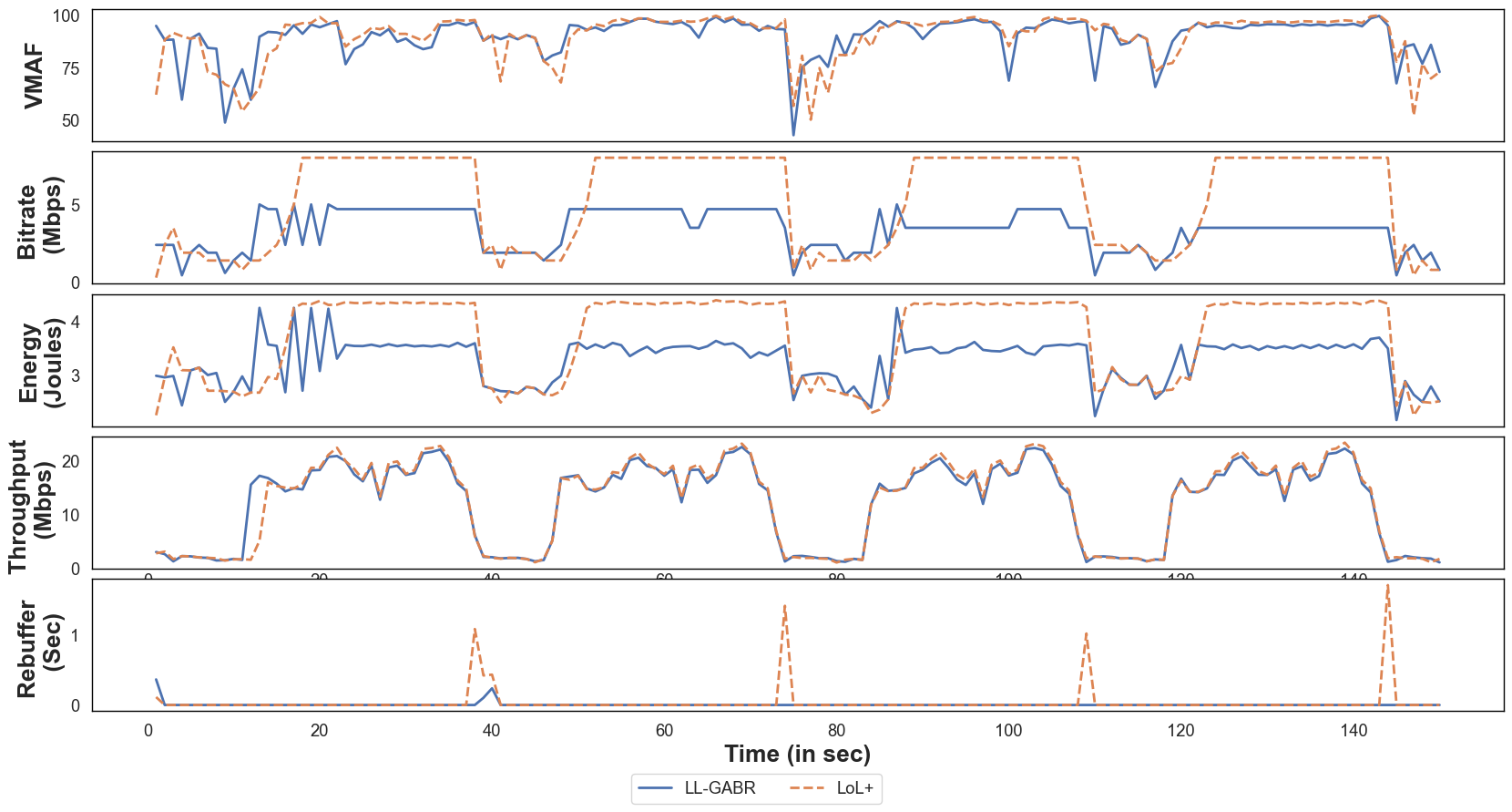}
\caption{Trace simulation for a fluctuating network.}
\label{fig:single_trace_1}
\end{figure}

In Figure \ref{fig:single_trace_1}, we have a network trace with a high degree of bandwidth fluctuation where we compare the performance of LoL+ and \ProjectName. We note that both ABRs are quick to adapt to the fluctuating network, dropping bitrate to manageable levels whenever there is a drop in bandwidth. However, it should be noted that LoL+ selects the highest bitrate when possible. Since LoL+ does not consider perceptual video quality when making bitrate decisions and since the bandwidth goes up to more than 20 Mbps which is more than sufficient to stream a video at an 8 Mbps bitrate, the ABR ends up selecting the highest bitrate. {\ProjectName} on the other hand takes into account both the VMAF values and energy consumption when making bitrate decisions. We, therefore, see that {\ProjectName} goes only up to 5 Mbps and is still able to achieve similar VMAF values to LoL+. This also results in a considerably smaller energy consumption in {\ProjectName} when compared to LoL+. We also see more freezing events in LoL+ as a consequence of selecting higher bitrates while {\ProjectName} has close to zero rebuffering events.



\subsubsection{\textbf{\textit{Training Performance:}}}
\label{ss:training_perf}
We train a distributed variant of the Soft-Actor-Critic network which significantly reduces training time and number of iterations to reach optimal reward values. In Table \ref{tab:cost}, we see how training time decreases when we increase the number of parallel workers and we also compare it against BDQ. BDQ takes close to 8 hours to converge rewards to an optimal value while SAC with 5 parallel workers takes less than an hour and reaches convergence in just 1500 episodes. This means it takes 5 independent agents each going through 1500 episodes working parallel to each other to collect experiences.

\subsubsection{\textbf{\textit{Energy \& Resource cost of {\ProjectName}}}}
\label{ss:cost}
We note the cost of both training and deploying the model in terms of energy consumption, CPU, and memory usage in Table \ref{tab:cost}

\textbf{Training Cost:}
We use Intel's RAPL interface to directly read the CPU energy consumption of the system while the model is training. We read the \textit{energy\_uj} file at every episode interval to find the energy in \textit{Joules} per episode. We also read the value for the CPU percentage and memory being used while training using the python \textit{psutil} library. We compare these values for different iterations of SAC and BDQ. We note that cost in terms of CPU\% and energy increases as the number of cores or parallel workers increases. BDQ on the other due to its much larger training time uses about 6 times as much energy as the SAC model. The CPU\% for SAC-5 and SAC-3 is divided among the cores that are used for each worker, which translates to an average of around 50\%/core and 60\%/core for SAC-5 \& SAC-3 respectively.

It should be noted that training is performed only once and the trained model can be used indefinitely on different videos. Table \ref{tab:all_results} shows us that {\ProjectName} saves us around \textit{0.1Kj} of energy for a 5-minute streaming session. If we consider an average viewing session as 120 minutes for a football match, this results in \textit{2.4Kj} of energy saved and it would take just 65 streaming sessions to offset the cost of training {\ProjectName} using \textit{SAC-5} (156Kj).

\textbf{Inference Cost:} The energy expenditure for executing our RL model on a mobile device was assessed by conducting model inferences in isolation from video streaming, with energy consumption monitored via Android's BatteryManager. We use \textit{Samsung Galaxy A10} and \textit{Motorolo One} devices for these experiments. Over a duration of five minutes, with inferences invoked every second, the model's energy usage was around 0.2 \textit{Joules} which is less than 0.01$\%$ of the streaming energy consumption. Given its minimal impact, this energy cost is negligible compared to the total energy required for streaming, rendering it inconsequential in the broader context of streaming energy expenditures.

%% file: mega_table_3.tex
\begin{table*}[ht]
\centering
\scriptsize
\begin{tabular}{c|c|cccccccccc}
     
\hline

& \multirow{2}{*}{\textbf{ABR}} & 
\multirow{1}{*}{\textbf{Quality}} & 
\multirow{1}{*}{\textbf{Quality}} & 
\multirow{1}{*}{\textbf{Data}} & 
\multirow{1}{*}{\textbf{Bitrate}} & 
\multirow{1}{*}{\textbf{Latency}} & 
\multirow{2}{*}{\textbf{Speed}} & 
\multirow{1}{*}{\textbf{Freezing}} & 
\multirow{1}{*}{\textbf{Energy}} & 
\multirow{2}{*}{\textbf{QoE} $\uparrow$} & 
\multirow{1}{*}{\textbf{Energy}} \\

&  & \textbf{level} (0-5) $\uparrow$ & \textbf{Smooth} $\downarrow$& \textbf{dl} \textit{(in Mb)} $\downarrow$ & (in Mbps) $\downarrow$ & \textit{(sec)} $\downarrow$ & & \textit{(sec)} $\downarrow$& \textit{(kj)} $\downarrow$& & \textbf{Efficiency} $\uparrow$ \\
\hline
\multirow{5}{*}{\rotatebox{90}{\textbf{Animation}}}
    & {\ProjectName} & 
         4.5{\scriptsize$\pm$0.18} & \cellcolor{green!20}{10.6\scriptsize{$\pm$12.2} }  & \cellcolor{green!20}{75.6\scriptsize{$\pm$19.7} }& \cellcolor{green!20}{3.02$\pm$0.81} & 2.75\scriptsize{$\pm$2.6} &1.01\scriptsize{$\pm$0.01} & \cellcolor{green!20}{3.36\scriptsize{$\pm$7.3} } & \cellcolor{green!20}{0.90\scriptsize{$\pm$0.08} } & \cellcolor{green!20}{69.7\scriptsize{$\pm$35.5} } & \cellcolor{green!20}{45.4\scriptsize{$\pm$22.8 }} \\
    & BDQ & 
        \cellcolor{green!20}{4.60{\scriptsize$\pm$0.18 } } & \cellcolor{red!20}{16.67\scriptsize{$\pm$15.7}} & \cellcolor{red!20}{165.04\scriptsize{$\pm$40.99} }& \cellcolor{red!20}{6.05$\pm$1.44} & \cellcolor{red!20}{5.47 \scriptsize{$\pm$2.1}}& \cellcolor{green!20}{1.00\scriptsize{$\pm$0.01}} & 3.77\scriptsize{$\pm$5.14}  & \cellcolor{red!20}{1.18\scriptsize{$\pm$0.10} } & 54.71\scriptsize{$\pm$42.75} & 28.99\scriptsize{$\pm$22.41} \\
    & LoL+ & 
        4.20{\scriptsize$\pm$0.63} & 12.52\scriptsize{$\pm$9.92} & 95.51\scriptsize{$\pm$65.0}& 3.37$\pm$2.36& \cellcolor{green!20}{1.98 \scriptsize{$\pm$1.8} } &1.01\scriptsize{$\pm$0.02} & 4.19\scriptsize{$\pm$7.93} &0.96\scriptsize{$\pm$0.21} & 51.19 \scriptsize{$\pm$46.19}& 30.36\scriptsize{$\pm$28.85} \\
    & Stallion & 
        \cellcolor{red!20}{3.97{\scriptsize$\pm$0.68}} & 14.26\scriptsize{$\pm$11.3} & 92.96\scriptsize{$\pm$56.24}&3.33$\pm$2.08 & 2.14\scriptsize{$\pm$1.5}  &\cellcolor{red!20}{1.02\scriptsize{$\pm$0.02}} & 5.01\scriptsize{$\pm$8.15} &0.96\scriptsize{$\pm$0.18} & \cellcolor{red!20}{44.05\scriptsize{$\pm$53.66}} & \cellcolor{red!20}{25.23\scriptsize{$\pm$34.29}}\\
    & L2A & 
        4.58{\scriptsize$\pm$0.14} & 11.67\scriptsize{$\pm$8.32} & 141.85\scriptsize{$\pm$54.24}& 5.12$\pm$1.93&2.78\scriptsize{$\pm$2.7}&1.01\scriptsize{$\pm$0.02} &\cellcolor{red!20}5.50\scriptsize{$\pm$10.74} & 1.12\scriptsize{$\pm$0.14}& 67.22\scriptsize{$\pm$47.31}& 38.37\scriptsize{$\pm$26.83}\\
    \hline
    
\multirow{5}{*}{\rotatebox{90}{\textbf{Movie}}} 
    & {\ProjectName} 
        & 3.61\scriptsize{$\pm$0.20}&4.74\scriptsize{$\pm$8.03} & \cellcolor{green!20}{74.49\scriptsize{$\pm$29.99} }&\cellcolor{green!20}{2.36$\pm$0.30}& 2.72\scriptsize{$\pm$2.14} &\cellcolor{green!20}{1.01\scriptsize{$\pm$0.02}} & 3.52\scriptsize{$\pm$7.68} & \cellcolor{green!20}{0.82 \scriptsize{$\pm$0.04}} & \cellcolor{green!20}{55.34\scriptsize{$\pm$43.79 }} & \cellcolor{green!20}{43.67\scriptsize{$\pm$26.07 }}\\
    & BDQ 
        & 4.10\scriptsize{$\pm$0.34}&\cellcolor{red!20}{40.66\scriptsize{$\pm$37.05}} & \cellcolor{red!20}{215.25\scriptsize{$\pm$84.02}}&\cellcolor{red!20}{5.83$\pm$2.28} & \cellcolor{red!20}{5.90\scriptsize{$\pm$3.22}} &\cellcolor{red!20}{0.96\scriptsize{$\pm$0.01}} & \cellcolor{green!20}{3.05\scriptsize{$\pm$4.21 }} & \cellcolor{red!20}{1.10\scriptsize{$\pm$0.12}} & \cellcolor{red!20}{9.25\scriptsize{$\pm$71.81}}& \cellcolor{red!20}{0.99\scriptsize{$\pm$44.39}}\\
    & LoL+ 
        & 3.50\scriptsize{$\pm$0.92}&5.18\scriptsize{$\pm$6.04} & 135.77\scriptsize{$\pm$107.00}&3.69$\pm$2.90& \cellcolor{green!20}{2.07\scriptsize{$\pm$2.15 }} &1.02\scriptsize{$\pm$0.02} & 4.95\scriptsize{$\pm$8.39}& 0.94\scriptsize{$\pm$0.18}& 39.46\scriptsize{$\pm$48.54}& 23.45\scriptsize{$\pm$30.88}\\
    & Stallion 
        & \cellcolor{red!20}{3.26\scriptsize{$\pm$0.93}}&\cellcolor{green!20}{3.96\scriptsize{$\pm$5.58 }} & 135.56\scriptsize{$\pm$96.59}&3.67$\pm$2.61& 2.27\scriptsize{$\pm$1.71}&1.02\scriptsize{$\pm$0.02} &5.69\scriptsize{$\pm$8.74}& 0.94\scriptsize{$\pm$0.16}& 39.23\scriptsize{$\pm$48.35} & 22.68\scriptsize{$\pm$30.44}\\
    & L2A 
        & \cellcolor{green!20}{4.10\scriptsize{$\pm$0.32} }&14.20\scriptsize{$\pm$11.67} & 208.93\scriptsize{$\pm$96.99}&5.70$\pm$2.62& 4.09\scriptsize{$\pm$5.97}&1.02\scriptsize{$\pm$0.03}& \cellcolor{red!20}{9.00\scriptsize{$\pm$18.53}} & 1.07\scriptsize{$\pm$0.15}& 40.31\scriptsize{$\pm$74.49}& 23.31\scriptsize{$\pm$42.66}\\
    \hline

\multirow{5}{*}{\rotatebox{90}{\textbf{Sports}}} 
    & {\ProjectName} 
        & 3.99\scriptsize{$\pm$0.12}&5.47\scriptsize{$\pm$7.88} &\cellcolor{green!20}{104.90\scriptsize{$\pm$13.89 }} &\cellcolor{green!20}{3.40$\pm$0.43} &3.15\scriptsize{$\pm$2.58}&\cellcolor{green!20}{1.01\scriptsize{$\pm$0.02}} &4.06\scriptsize{$\pm$8.38} & \cellcolor{green!20}{0.89\scriptsize{$\pm$0.02} } & \cellcolor{green!20}{ 65.19\scriptsize{$\pm$38.42} } &\cellcolor{green!20}{ 41.78\scriptsize{$\pm$24.46 }}\\
    & BDQ 
        & 3.77\scriptsize{$\pm$0.55}&\cellcolor{red!20}{38.27\scriptsize{$\pm$35.83}} & 136.82\scriptsize{$\pm$28.82}& 4.19$\pm$0.89&\cellcolor{red!20}{18.97\scriptsize{$\pm$1.13}}&\cellcolor{red!20}{0.90\scriptsize{$\pm$0.00}} &\cellcolor{green!20}{0.51\scriptsize{$\pm$1.51 }}&  0.98\scriptsize{$\pm$0.06}& \cellcolor{red!20}{4.33\scriptsize{$\pm$64.28}} & \cellcolor{red!20}{1.24\scriptsize{$\pm$39.36}}\\
    & LoL+ & 
        3.53\scriptsize{$\pm$0.86}&6.65\scriptsize{$\pm$7.64} & 128.67\scriptsize{$\pm$109.90}&3.96$\pm$3.40& \cellcolor{green!20}{2.00\scriptsize{$\pm$2.02}} &1.01\scriptsize{$\pm$0.02} & 4.14\scriptsize{$\pm$8.43}& 0.92\scriptsize{$\pm$0.18}& 48.51\scriptsize{$\pm$53.90}& 29.31\scriptsize{$\pm$35.24} \\
    & Stallion 
        & \cellcolor{red!20}{3.37\scriptsize{$\pm$0.85}} &\cellcolor{green!20}{5.39\scriptsize{$\pm$8.10 }} & 130.56\scriptsize{$\pm$101.76}&4.02$\pm$3.15&2.26\scriptsize{$\pm$1.72}& 1.02\scriptsize{$\pm$0.02}&5.06\scriptsize{$\pm$9.17}& 0.92\scriptsize{$\pm$0.16}& 49.94\scriptsize{$\pm$54.14}& 29.94\scriptsize{$\pm$35.15} \\
    & L2A 
        & \cellcolor{green!20}{4.15\scriptsize{$\pm$0.29}} &9.69\scriptsize{$\pm$10.46} & \cellcolor{red!20}{205.03\scriptsize{$\pm$103.68}}&\cellcolor{red!20}{6.35$\pm$3.21} & 3.78\scriptsize{$\pm$5.21}&  1.02\scriptsize{$\pm$0.03}&\cellcolor{red!20}{ 8.12\scriptsize{$\pm$16.40} } & \cellcolor{red!20}{1.05\scriptsize{$\pm$0.16}} & 51.15\scriptsize{$\pm$68.21}  & 30.20\scriptsize{$\pm$39.63} \\
    \hline

\end{tabular}
\caption{Average Results of all metrics across testing traces for different video datasets}
\label{tab:all_results}
\end{table*}

%% file: conclusion.tex
In this paper, we present {\ProjectName}, an RL-based ABR algorithm for low-latency video streaming on battery-constrained devices that maximizes the perceptual video quality while simultaneously minimizing total energy consumption. We utilize a Distributed Soft-Actor-Critic model to train our RL agent to maximize the user's QoE. Our experiments using real-world and synthetic network traces show that {\ProjectName} is able to outperform the SOTA benchmarks with a 35$\% (\downarrow)$ decrease in data usage, 44$\% (\uparrow)$ increase in QoE, 11$\% (\downarrow)$ savings energy consumption and 73$\% (\uparrow)$ increase in energy efficiency. {\ProjectName} is able to achieve these gains by maintaining high levels of video quality which is measured by VMAF and low levels of latency \& freezing values which are similar to current state-of-the-art models.

For future work, we will look to apply the {\ProjectName} on the server side to and build a multi-agent ABR to manage energy consumption across multiple devices operating on the same wireless network. We also hope to investigate the impact of different video codes like H.265 and AV1 on mobile energy consumption.